\documentclass[12pt]{iopart}
\usepackage{graphicx,epsfig}
\usepackage{amsthm}
\usepackage{color}

\begin{document}

\title[]{Interconversion between block coherence and multipartite entanglement in many-body systems}

\author{Yu-Hui Wang$^{1}$, Li-Hang Ren$^{1*}$, Ming-Liang Hu$^{2\dagger}$ and Yan-Kui Bai$^{1\ddagger}$}

\address{$^{1}$College of Physics and Hebei Key Laboratory of Photophysics Research and Application, Hebei Normal University, Shijiazhuang, Hebei 050024, China}
\address{$^{2}$School of Science, Xi’an University of Posts and Telecommunications, Xi’an 710121, China}

\ead{renlihang@hebtu.edu.cn}
\ead{mingliang0301@163.com}
\ead{ykbai@semi.ac.cn}
\vspace{10pt}


\begin{abstract}
Coherence is intrinsically related to projective measurement. When the fixed projective measurement involves higher-rank projectors, the coherence resource is referred  to as block coherence, which comes from the superposition of orthogonal subspaces. Here, we establish a set of quantitative relations for the interconversion between block coherence and multipartite entanglement under the framework of the block-incoherent operations. It is found that the converted  multipartite entanglement is upper bounded by the initial block coherence of single-party system. Moreover, the generated multipartite entanglement can be transferred to its subsystems and restored to block coherence of the initial single-party system by means of local block-incoherent operations and classical communication. In addition, when only the coarse-grained quantum operations are accessible for the ancillary subsystems, we further demonstrate that a lossless resource interconversion is still realizable, and give a concrete example in  three four-level systems. Our results provide a versatile approach to utilize different quantum resources in a cyclic fashion.
\end{abstract}

%
%
%
%
%

\section{Introduction}
Both quantum coherence and entanglement are crucial physical resources in quantum information processing \cite{coherence,hu,Horodecki09rmp}. It was shown that quantum coherence and entanglement can be interconverted in bipartite and multipartite systems under certain conditions, which provides an operational connection between these two kinds of quantum resources \cite{Streltsov15prl,chitambar16prl,zhu17pra,zhu18pra,Qi17,ren}. Moreover, operational methods for other resource conversions concerning nonclassicality, quantum correlation, and nonlocality were also put forward \cite{Killoran16prl,ma16prl,Tan19pra,XiY19pra}, and experimental explorations have been demonstrated in the optical and superconducting systems \cite{wuoptica,experiment,qiao18pra,superconducting}.

In general, quantum coherence is based on a fixed orthonormal basis $\{|i\rangle\}$. The standard resource theory of quantum coherence was constructed by Baumgratz \textit{et} \textit{al}  \cite{Baumgratz14prl}, in which the states that are diagonal in the fixed basis $\{|i\rangle\}$ are incoherent while states that do not conform to this  form are coherent.  In other words, the  incoherent states can be obtained by a dephasing operation consisting of projectors $\{|i\rangle\langle i|\}$, which corresponds to a rank-1 projective measurement and can be regarded as the fine-grained projective measurement. When the fine-grained projective measurement is unavailable, the coherence exhibits in the form of block coherence, which comes from the superposition of orthogonal subspaces spanned by  higher-rank projectors that correspond to a coarse-grained projective measurement \cite{aberg,brub19prl,brub21pra}. The resource theory of block coherence plays an important role in characterizing resource states, where the observers cannot perform the fine-grained projective measurements. For example, it is the case that the observers can only estimate whether the spins of two spin-1/2 particles are parallel or antiparallel \cite{Gisin99prl,Massar95prl,Bartlett04pra,Rezazadeh17pra,Rezazadeh19pra,Mani2024pra}.

It is desirable to explore the operational connection between block coherence and entanglement from the viewpoint of experimental operations, although the resource conversion between quantum coherence and multipartite entanglement was investigated \cite{ren}. Recently, an operational method was proposed to convert block coherence to bipartite entanglement via a bipartite block-incoherent operation \cite{kim21pra}. However, under the framework of full block-incoherent scenario, it remains an open problem whether block coherence and quantum entanglement can be interconverted, especially for the case of multipartite entanglement due to it being a precious resource in multi-party quantum information processing. Moreover, when only the coarse-grained quantum operations are accessible, it is necessary to find the optimal operations which can realize the cyclic conversion between block coherence and multipartite entanglement without the loss.

In this paper, we first briefly review the resource theory of multipartite block coherence, and then explore the interconversion between block coherence and multipartite entanglement under the framework of full block-incoherent operations. It is shown that block coherence of the initial single-party system can be converted to multipartite entanglement via multipartite block-incoherent operations, where we establish a rigorously quantitative relation. In the reversed process, multipartite entanglement can be cyclically converted to block coherence of local subsystems by utilizing local block-incoherent operations and classical communication (LBICC). Finally, when only the coarse-grained projective measurements can be performed on the ancillary subsystems, we further demonstrate that a lossless resource interconversion is still realizable.

\section{Resource theory of multipartite block coherence}
It is known that quantum coherence is based on a fixed orthogonal basis, which can be viewed as a rank-1 projective measurement. From this point, {\AA}berg introduced the measure to quantify superposition with respect to general projective measurement whose projector may have an arbitrary rank \cite{aberg}, which was later termed as block coherence \cite{brub19prl}. Considering a general projective measurement $\textbf{P}=\{P_{i}\}$, where the rank of every projector $P_{i}$ is arbitrary, the block-incoherent states are defined as \cite{aberg,brub19prl}
\begin{equation}\label{BIstate}
	\rho_{BI}=\sum_i P_i \rho P_i=\bigtriangleup[\rho],\quad\rho\in\mathcal{S},
\end{equation}
where	 $\mathcal{S}$ is the set of quantum states and $\bigtriangleup$ represents the block-dephasing operation. Denoting $\mathcal{I}_{BI}$ as the set of all block-incoherent quantum states,   the block-incoherent operation $\Lambda_{BI}$  is a  channel that maps  any block-incoherent state to another block-incoherent state, i.e., $\Lambda_{BI}(\rho_{BI})\subseteq\mathcal{I}_{BI}$.  A quantum channel is usually expressed by Kraus operators, so the block-incoherent operation can be written as $\Lambda_{BI}(\rho)=\sum_l K_l\rho K_l^{\dagger}$ with $\{K_l\}$ satisfying $K_l\mathcal{I}_{BI}K_l^{\dagger}\subseteq\mathcal{I}_{BI}$ and $\sum_l K_l^{\dagger}K_l=\textbf{I}$ \cite{brub21pra}. In analogy to the case of standard coherence theory, block-incoherent Kraus operator have a similar form, which reads \cite{brub21pra}
\begin{equation}
	K_{l}=\sum_{i}P_{f_{l}(i)}C_{l}P_{i},
\end{equation}
where the subscript $f_{l}(i)$ is some index function, and $C_{l}$ is a complex matrix satisfying the normalization condition. Based on these, the block coherence can be quantified by suitable measures \cite{aberg,brub19prl,brub21pra,fei20pra,Yan2022,ma23prr,Luo22pra,Bosyk21pra,Srivastava21pra}. Here, we focus on the relative entropy of block coherence, which has the form \cite{aberg,brub19prl}
\begin{equation}\label{Cr}
	C_{R}(\rho;\textbf{P})=\min_{\sigma\in\mathcal{I}_{BI}}S(\rho\|\sigma)=S(\bigtriangleup[\rho])-S(\rho),
\end{equation}
where $S(\rho\|\sigma)=\tr(\rho\log_2\rho-\rho\log_2\sigma)$ is the quantum relative entropy, and $S\left(\rho\right)=-\tr(\rho\log_{2}\rho)$ is the von Neumann entropy. Note that the concepts mentioned above coincide with their counterparts in the standard resource theory of coherence when all the projectors are rank-1 cases.

Similar to the standard  resource theory of multipartite coherence \cite{sun15pra}, the framework of block coherence can also be generalized to multipartite systems. The bipartite block coherence was discussed in Ref. \cite{fei20pra}, and we further consider the case of an $N$-partite system. By choosing the fixed projectors to be $\textbf{P}_{N}=\{P_{i_1}^{A_1}\otimes P_{i_2}^{A_2}\otimes\cdots\otimes P_{i_n}^{A_n}\}$, an $N$-partite block-incoherent states can be defined as
\begin{equation}\label{BIm}
	\rho^N_{BI}=\sum_s p_s \sigma_s^{A_1}\otimes \sigma_s^{A_2}\otimes\cdots \otimes\sigma_s^{A_n},
\end{equation}
where $p_s$ are probabilities, $\sigma_s^{A_1}$ is a block-incoherent state on the subsystem $A_1$, i.e.,$\sigma_s^{A_1}=\sum_{i_1}P_{i_1}^{A_1}\rho_s^{A_1}P_{i_1}^{A_1}$ with $\rho_s^{A_1}$ being any state in the Hilbert space  of subsystem $A_1$, and the situation of $\sigma_s^{A_k} (k=2,\cdots,n)$ is similar. When we use $\mathcal{I}_{BI}^N$ to represent the set of all $N$-partite block-incoherent states, the $N$-partite block-incoherent operations can still be written as Kraus operators $\{K_l\}$, where the operators map every $N$-partite block-incoherent state to some other one, i.e., $K_l\mathcal{I}^N_{BI}K_l^{\dagger}\subseteq\mathcal{I}^N_{BI}$. It is worth noting that multipartite block coherence can also be quantified by the relative entropy of block coherence in Eq. (\ref{Cr})  with respect to $\textbf{P}_N=\{P_{i_1}^{A_1}\otimes P_{i_2}^{A_2}\otimes\cdots\otimes P_{i_n}^{A_n}\}$.

\section{Resource conversion from block coherence to multipartite entanglement}

In this section, we study the resource conversion from block coherence to multipartite entanglement via multipartite block-incoherent operations. In comparison to bipartite entanglement, multipartite entanglement can characterize some special tasks in multi-party systems, such as multipartite entanglement dynamics \cite{meyer02jmp,car04prl,bai09pra,bai14prl,chi18rpp}, quantum phase transitions \cite{ost02nat,oli06prl,hof14prb,bai22pra,li24njp}, and so on. On the other hand, we noted that multipartite block-incoherent operations are not equivalent to those of the bipartite case in general, since the multipartite operations have the ability to generate multipartite entangled states (see the details in appendix A).

Here we consider the relative entropy of block coherence given in Eq. (3), and accordingly the relative entropy of multipartite entanglement is adopted. For an $N$-partite quantum state $\rho_{A_{1}\cdots A_{n}}$, the multipartite relative entropy of entanglement is defined as \cite{Modi10prl,Vedral1997}
\begin{equation}\label{Er}
	E_{R}(\rho_{A_{1}\cdots A_{n}})=\min_{\delta_{A_{1}\cdots A_{n}}\in \mathcal{D}}S(\rho_{A_{1}\cdots A_{n}}\|\delta_{A_{1}\cdots A_{n}}),
\end{equation}
where $\delta_{A_{1}\cdots A_{n}}$ is the $N$-partite fully separable state, and $\mathcal{D}$   denotes the set of all  fully separable states.  Here we use $\Lambda_{BI}^m$ to represent a multipartite block-incoherent operation. The block coherence of a single-party system $A$ can be converted to multipartite entanglement by attaching ancillas $B_1B_2\cdots B_n$ and then applying a multipartite block-incoherent operation $\Lambda_{BI}^m$. The quantitative relation in this process goes as follows.

\textbf{Theorem 1.} Applying a multipartite block-incoherent operation $\Lambda_{BI}^m$  to a  block-coherent state $\rho _{A}$ and the ancillary $N$-partite block-incoherent state $\sigma _{B_{1}B_{2} \dots B_{n}}$, the generated multipartite relative entropy of entanglement is upper bounded by the relative entropy of block coherence in $\rho _{A}$, namely,
\begin{equation}\label{theorem1}
    E_{R} [\Lambda_{BI}^m(\rho_A\otimes\sigma _{B_{1}B_{2}\cdots B_{n}}) ]\leq C_{R} (\rho _{A};\textbf{P}),
\end{equation}
where $\textbf{P}=\{P^A_{i}\}$,  $\Lambda_{BI}^m$ is an $(N+1)$-partite block-incoherent operation with respect to general projective measurement $\{P_i^A\otimes P^{B_1}_{j_1}\otimes P^{B_2}_{j_2}\otimes\cdots\otimes P^{B_n}_{j_n}\}$, and  $E_R$ is the $(N+1)$-partite relative entropy of entanglement.

\textbf{Proof}.---
Letting $\sigma_{A}$ be  the closest block-incoherent state to $\rho_{A}$, then according to the definition of relative entropy of block coherence,  we have
\begin{eqnarray}
C_{R}(\rho_{A};\textbf{P})&=&S(\rho_{A}\|\sigma_{A}) \nonumber\\
&=&S\left ( \rho _{A}\otimes \sigma _{B_{1}B_{2} \dots B_{n}} \parallel \sigma _{A}\otimes \sigma _{B_{1}B_{2} \dots B_{n}}  \right )\nonumber\\
&\ge& S\left [ \Lambda_{BI}^m \left ( \rho _{A}\otimes \sigma _{B_{1}B_{2} \dots B_{n}} \right )\parallel \Lambda_{BI}^m\left ( \sigma _{A} \otimes \sigma _{B_{1}B_{2} \dots B_{n}} \right ) \right ]\nonumber\\
&\ge& E_{R}\left [ \Lambda_{BI}^m\left ( \rho _{A}\otimes \sigma _{B_{1}B_{2} \dots B_{n}}\right ) \right ],
\end{eqnarray}
where the additive and contractive properties of relative entropy are used in the second  and  third lines, and the result of the last inequality comes from the definition of the relative entropy of entanglement.
\textcolor{black}{\qed}

It is noted that the quantitative relation, analogous to Eq. (\ref{theorem1}),  also holds for the  block coherence and  bipartite entanglement for arbitrary bipartition  in many-body systems $AB_1B_2\cdots B_n$ (the details are presented in Appendix A).

Here we want to find a multipartite block-incoherent operation $\Lambda_{BI}^m$ to saturate the relation in Eq. (\ref{theorem1}).  Since the fixed  projective measurement and ancillary states can be arbitrarily chosen, we consider a special case, in which the fixed projectors are chosen to be $\{P_{i}\otimes|j_{1}\rangle\langle j_{1}|\otimes\cdots\otimes|j_{n}\rangle\langle j_{n}|\} $, and the ancillary states are selected as $|00\cdots0\rangle\langle00\cdots0|_{B_{1}B_{2}\cdots B_{n}}$.   For convenience, we have omitted the superscripts of projectors. Assuming the number of projectors in every subsystem is $d$,  the inequality in Eq. (\ref{theorem1})  reaches saturation when we apply the following multipartite block-incoherent unitary operator
\begin{equation}\label{U}
	U_m=\sum_{i,j_{1},\cdots, j_{n}=0}^{d-1}P_{i}\otimes|mod(i+j_{1},d)\rangle\langle j_{1}|\otimes\cdots\otimes|mod(i+j_{n},d)\rangle\langle j_{n}|.
\end{equation}
The generated multipartite quantum state reads
\begin{eqnarray}
\rho_{m}&=&U_m\left(\rho_{A}\otimes \left|00\cdots0\right\rangle\left\langle00\cdots0\right|\right)U_m^{\dagger}\nonumber\\
&=&\sum_{i,j=0}^{d-1}P_{i}\rho_{A}P_{j}\otimes\left|ii\cdots i\right\rangle\left\langle jj\cdots j\right|_{B_{1}B_{2}\cdots B_{n}},
\end{eqnarray}
which is the optimal generated state in this scenario. In fact, $\rho_m$ is a  multipartite entangled state \cite{dur99prl,huber14prl,Vicente11pra,dai20prap,Szalay15pra}, which has fundamental  difference from the bipartite case in Ref. \cite{kim21pra} (see Appendix A for detail).

In the following, we will prove that this process is an optimal conversion, i.e.,
$E_{R} ( \rho _{m}   )=C_{R}( \rho _{A};\textbf{P} )$. Denoting $\{|k^{(i)}\rangle\}_{k}$ as a basis of the subspace spanned by the range of $P_{i}$, the matrix element of $\rho_A$ given by bases $|k^{ ( i  ) }\rangle$ and $|l^{(j)} \rangle$ is
\begin{eqnarray}
	\rho_{k^{(i)}l^{(j)}}&=& \langle k^{ ( i  ) }   | \rho _{A}   | l^{(j) }   \rangle= \langle k^{ (i) }   |P_{i} \rho _{A}P_{j}  | l^{(j)}   \rangle\nonumber\\
	&=& \langle k^{ ( i  ) }ii\cdots i  |\rho_{m}  | l^{(j)}jj\cdots j  \rangle,
\end{eqnarray}
which is embedded in  $\rho_{m}$. Therefore, we conclude  that the non-zero matrix elements of  $\rho_{m}$ are the same as those of $\rho_{A}$, and the other matrix elements of $\rho_{m}$ are zero. This implies that $S\left(\rho_{m}\right)=S\left(\rho_{A}\right)$. Furthermore, the reduced state of subsystem $A$ in $\rho_m$ is obtained by $\rho_{A}'=\tr_{_{B_{1}B_{2}\cdots B_{n}}}\rho_m=\sum_iP_i \rho_A P_i=\triangle\left(\rho_{A}\right)$.  In this paper, we label $E_R$ as the  multipartite entanglement, and now we define $E_R^{A|B_{1}B_{2}\cdots B_{n}}$ as the bipartite relative entropy of entanglement in the partition $A|B_{1}B_{2}\cdots B_{n}$.  Since multipartite relative entropy of entanglement is not smaller than bipartite relative entropy of entanglement in an arbitrary bipartition \cite {ren}, we have
\begin{eqnarray}
E_{R}(\rho_{m})&\geq&E_R^{A|B_{1}B_{2}\cdots B_{n}}(\rho_m) \geq S(\rho_{A}')-S(\rho_m) \nonumber\\
&=& S\left[\bigtriangleup\left(\rho_{A}\right)\right]-S\left(\rho_{A}\right)=C_{R}\left(\rho_{A};\textbf{P}\right),
\end{eqnarray}
where in the second inequality we have used the relation $E^{A|B}_{R}\left(\rho_{AB}\right)\geq S\left(\rho_{A}\right)-S\left(\rho_{AB}\right)$ \cite{Plenio}. According to  Theorem 1, we have $E_{R}(\rho_{m})\leq C_{R}(\rho_{A};\textbf{P})$, so  $E_{R}(\rho_{m})=C_{R}(\rho_{A};\textbf{P})$ is proved.  That is to say, $U_m$ given in Eq. (8) is an optimal block-incoherent operation in the resource conversion from block coherence to multipartite entanglement.

\section{Resource conversion from multipartite entanglement to block coherence}
It has been shown that block coherence can be converted to multipartite entanglement in above section. Next, we further explore the reversed process. Since  $\rho_m$ is an optimal output state that  acquires the same amount of multipartite entanglement as that of initial block coherence, the reversed process should start from the multipartite state $\rho_m$, in order to  complete  a cyclic scheme. Before exploring the resource conversion from multipartite entanglement to block coherence, we give a relation between entanglement and block coherence for state $\rho_m$.

\textbf{Theorem 2.} For the  generated  state $\rho_{m}$ obtained by applying the optimal multipartite block-incoherent operation $U_m$ to initial state $\rho_A$ and its ancillas $|0\rangle\langle 0|^{\otimes n}$,  the following relation holds
\begin{eqnarray}
 E_{R}(\rho_{m})=C_{R}(\rho_{A};\textbf{P})=C_{R}(\rho_{m};\textbf{P}_m),
\end{eqnarray}
where  $\rho_{m}=\sum_{i,j}P_{i}\rho_{A}P_{j}\otimes\left|ii\cdots i\right\rangle\left\langle jj\cdots j\right|_{B_{1}B_{2}\cdots B_{n}}$, the projectors $\textbf{P}=\{P_{i}\}$
and $\textbf{P}_m=\{P_{i}\otimes|j_{1}\rangle\langle j_{1}|\otimes\cdots\otimes|j_{n}\rangle\langle j_{n}|\}$.

\textbf{Proof}.---
According to the definition of multipartite relative entropy of block coherence, we have $C_{R}(\rho_m;\textbf{P}_m)=S[\bigtriangleup(\rho_{m})]-S(\rho_m)$. The block-diagonal part of the state $\rho_{m}$ is
\begin{equation}
	\bigtriangleup(\rho_{m})=\sum_{i=0}^{d-1}P_{i}\rho_{A}P_{i}\otimes|ii\cdots i \rangle \langle ii\cdots i |_{B_{1}B_{2}\cdots B_{n}},
\end{equation}
and the  state $\bigtriangleup(\rho_{A})$ only leaves   block-diagonal matrix elements, which gives
\begin{eqnarray}
	\rho_{k^{(i)}l^{(i)}}&=& \langle k^{(i)} |P_{i} \rho _{A}P_{i} | l^{({i})  }  \rangle\nonumber\\
	&=& \langle k^{(i)}ii\cdots i  |\bigtriangleup (\rho_{m}) | l^{ (i)  }ii\cdots i \rangle.
\end{eqnarray}
This means that the matrix elements of $\bigtriangleup(\rho_{A})$ are embedded in the matrix of $\bigtriangleup(\rho_{m})$, which gives $S[\bigtriangleup(\rho_{m})]=S[\bigtriangleup(\rho_{A})]$.  Since we have obtained $S(\rho_{m})=S(\rho_{A})$ in the above discussion,  thus $C_{R}(\rho_{m};\textbf{P}_m)=C_{R}(\rho_{A};\textbf{P})$. Because $\rho_m$ is an optimal output state in the conversion from  block coherence to multipartite entanglement, we can obtain  $E_{R}(\rho_{m})=C_{R}(\rho_{A};\textbf{P})=C_{R}(\rho_{m};\textbf{P}_m)$, which completes the proof.
\textcolor{black}{\qed}

In the following, we will study how to restore block coherence of local subsystems from multipartite entanglement. In standard resource theory of coherence, this task was first introduced in bipartite systems, which was referred to as the assisted distillation of quantum coherence  by a class of local quantum incoherent operations and classical communication (LQICC) \cite{Chitambar16prl}. This class of operations mean that one party performs arbitrary local quantum operations on its subsystem, while another one is restricted to local incoherent operations assisted by classical communication between them.  Another class, which was called local incoherent operations and classical communication (LICC),  was  proposed to set further limitations that all local operations on both parties should be incoherent  \cite{Strelsov17prx}.  Later on,  the LICC was applied in a cyclic resource conversion of coherence-entanglement-coherence, since it is free within the whole scenario \cite{ren}. Motivated by these, it is desirable to  propose a set of  local block-incoherent operations and classical communication (LBICC) in the multi-party systems, where all the parties  can only perform local block-incoherent operations and   communicate classically with each other. Using $\phi_{LBICC}$ to mark the  LBICC operation, our result is as follows.

\textbf{Theorem 3.} For the optimal output state under multipartite block-incoherent operation $U_m$, its multipartite relative entropy of entanglement is an upper bound on the block coherence of the reduced state transformed via LBICC
\begin{equation}
 C_{R}\left(\rho _{\alpha}^{LBICC};\textbf{P}_{\alpha} \right )\leq E_{R}(\rho_{m}),
\end{equation}
where $\rho _{\alpha}^{LBICC}=\tr_{\overline{\alpha}}\left[\phi_{LBICC}(\rho_{m})\right]$ is the reduced state of multipartite state $\phi_{LBICC}(\rho_{m})$ with $\overline{\alpha}$ being the traced subsystems, and $\textbf{P}_{\alpha}$ corresponds to the fixed projectors of the  remaining subsystems.

\textbf{Proof}.---  According to Theorem 2, we have the equation  $E_{R}\left(\rho_{m}\right)=C_{R}\left(\rho_{m};\textbf{P}_m\right)$ for the optimal output state $\rho_{m}$. Due to the property that  relative entropy  is not increasing after tracing  some subsystems out \cite{Vedral02rmp}, i.e.,
\begin{equation}\label{partialtrace}
S\left(\tr_{\overline{\alpha}}\rho\|\tr_{\overline{\alpha}}\sigma\right)\leq S\left(\rho\|\sigma\right),
\end{equation}
where $\tr_{\overline{\alpha}}$ is a partial trace, we  obtain
\begin{eqnarray}
E_{R}\left(\rho_{m}\right)&=&C_{R}\left(\rho_{m};\textbf{P}_m\right)\nonumber\\
&\geq& C_{R}\left[\phi_{LBICC}\left(\rho_{m}\right);\textbf{P}_m\right]\nonumber\\
&=&S\left[\phi_{LBICC}\left(\rho_{m}\right)\|\sigma\right]\nonumber\\
&\geq&S\left(\tr_{\overline{\alpha}}\left[\phi_{LBICC}\left(\rho_{m}\right)\right]\|\tr_{\overline{\alpha}}\sigma\right)\nonumber\\
&\geq&C_{R}\left(\rho _{\alpha}^{LBICC};\textbf{P}_{\alpha}\right),
\end{eqnarray}
where the first inequality is satisfied due to $C_R$ being monotone under the LBICC, in the third line $\sigma$ is the nearest block-incoherent state to $\phi_{LBICC}\left(\rho_{m}\right)$, and the last inequality follows from the definition of relative entropy of block coherence. Then the proof is completed.
\textcolor{black}{\qed}

Next, we will present the optimal LBICC operation to satisfy $E_{R}\left(\rho_{m}\right)=C_{R}\left(\rho _{\alpha}^{LBICC};\textbf{P}_{\alpha}\right)$. In this case, the local block-incoherent operation   can be chosen as $K_{l}=\frac{1}{\sqrt{d}}\sum_{k}e^{-i\phi _{k}^{l} } |l\rangle\langle k|$. Firstly, the block-incoherent measurement $\{K_{l} \}$ is performed on  subsystem $B_{n}$, and then the post-measurement state of $AB_{1}B_{2}\cdots B_{n-1}$ can be obtained from the $(N+1)$-partite  state to the $N$-partite  state via tracing $B_{n}$ out.  Then   a block-incoherent unitary operation $U_l=\sum_{k}e^{i\phi _{k}^{l} } |k\rangle\langle k|$ will be made on subsystem $B_{n-1}$ according to the measurement outcome $l$. Thus, the remaining state reads
\begin{equation}
	\rho_m^{\prime}=\sum_{i,j} P_{i}\rho_{A}P_{j}\otimes\left|ii\cdots i\right\rangle\left\langle jj\cdots j\right|_{B_{1}B_{2}\cdots B_{n-1}},
\end{equation}
which has a similar form to the  optimal state $\rho_m$ except that $\rho_m^{\prime}$ is an $N$-partite quantum state. Because $\rho_m$ and $\rho_m^{\prime}$ have the same nonzero matrix elements,
  entanglement and block coherence are transferred to the subsystems $AB_1\cdots B_{n-1}$ and keep the same amount, namely,
\begin{equation}\label{prime}
	 E_{R}(\rho_{m})=E_{R}(\rho_{m}^{\prime})=C_{R}(\rho_{m}^{\prime};\textbf{P}_m^{\prime})=C_{R}(\rho_{A};\textbf{P}),
\end{equation}
where $\textbf{P}_m^{\prime}=\{P_{i}\otimes|j_{1}\rangle\langle j_{1}|\otimes\cdots\otimes|j_{n-1}\rangle\langle j_{n-1}|\}$. Repeating the block-incoherent measurement and  block-incoherent unitary operation on all the subsystems $B_{i}$, a relation similar to Eq. (\ref{prime}) always holds in every step. Finally, the quantum state of the remained subsystem $A$ becomes $\rho_{A}^{f}=\sum_{i,j}P_{i}\rho_{A}P_{j}$ via $U^A_l=\sum_{k}e^{i\phi _{k}^{l} } P_{k}$, and thus the block coherence is restored to the single-party subsystem which satisfies  $C_{R}\left(\rho_{A}^{f};\textbf{P}\right)=E_{R}\left(\rho_{m}\right)=C_{R}(\rho _A;\textbf{P})$. Therefore, we realize the optimal resource conversion from multipartite entanglement to block coherence.

\section{Cyclic resource conversion in the scenario of full coarse-grained quantum operations}

In the previous two sections, we have demonstrated that block coherence and multipartite entanglement can be interconverted in multi-party systems, where it is assumed that the fine-grained projective measurements on the ancillary systems are accessible. However, when only the coarse-grained quantum operations are accessible, it remains an open problem whether the interconversion between block coherence and multipartite entanglement is realizable. In this section, we further explore the cyclic resource conversion in the scenario of full coarse-grained quantum operations.

Assume that the expression of fixed projectors is  $\textbf{P}_r=\left\{P_{i}\otimes P_{j_{1}}\otimes\cdots\otimes P_{j_{n}}\right\}$, and  each projector  has the same rank $r$.  For example, a  basis of subspace given by  the projector $P_{j_1}$ can be expressed as $\{|k^{(j_1)}\rangle\}_{k}$, and thus this projector  can be written as $P_{j_1}=\sum_{k=0}^{r-1}|k^{(j_1)}\rangle\langle k^{(j_1)}|$. So do the other projectors.  For an initial state $\rho_A$, letting the state of the ancillary systems $B_1B_2\cdots B_n$ be $|0^{(0)}\rangle\langle0^{(0)}|^{\otimes n}$, we propose an optimal multipartite block-incoherent operation that can convert  block coherence to multipartite entanglement, which has the following form
\begin{equation}
U_m^r=\sum_{i,j_{1},\cdots j_{n}=0}^{d-1}P_{i}\otimes\ P_{mod (i+j_{1},d)}C_{ij_{1}}P_{j_{1}}\otimes\cdots\otimes\ P_{mod (i+j_{n},d )}C_{ij_{n}}P_{j_{n}},
\end{equation}
with
\begin{equation}
C_{ij}=\sum_{n=0}^{r-1} |n^{mod (i+j,d )} \rangle \langle n^{ (j )} |.
\end{equation}
In Appendix B, we prove that this operation $U_m^r$ is a multipartite block-incoherent operation, and the matrices $\{C_{ij}\}$ make the normalization condition satisfied and realize the permutation of projectors.  In this scheme,
the generated state  is
\begin{equation}\label{rhomr}
	\varrho_{m}=
	\sum_{i,j=0}^{d-1} P_{i}\rho_{A}P_{j}\otimes |0^{(i)}0^{(i)}\cdots 0^{(i)} \rangle \langle 0^{(j)}0^{(j)}\cdots0^{(j)} |_{B_1B_2\cdots B_n},
\end{equation}
which is also a multipartite entangled state (see the details in the last paragraph of appendix B).  Now we verify that $U_m^r$ is an optimal multipartite block-incoherent operation. Firstly, it follows from Theorem 1 that
\begin{equation}\label{S}
	E_{R} (\varrho_{m} )\leq C_{R} (\rho_{A};\textbf{P} ).
\end{equation}
Since the matrix elements of $\rho_{A}$ are embedded in the matrix of $\varrho_{m}$ as follows:
\begin{eqnarray}
\rho_{k^{(i)}l^{(j)}}&=&\langle k^{ ( i ) } |\rho _{A} | l^{ ( {j} )  } \rangle
= \langle k^{ ( i ) } |P_{i}\rho _{A}P_{j} | l^{ ( {j} ) } \rangle\nonumber\\
&=& \langle k^{ ( i ) }0^{(i)}0^{(i)}\cdots 0^{(i)} |\varrho_{m} | l^{(j)}0^{(j)}0^{(j)}\cdots 0^{(j)} \rangle,
\end{eqnarray}
it means that $S (\varrho_{m} )=S (\rho_{A})$.  In addition, the reduced state of $A$ in $\varrho_{m}$ is obtained by 
$\rho_{A}^{\prime\prime}=\tr_{_{B_{1}B_{2}\cdots B_{n}}}\varrho_{m}=\bigtriangleup (\rho_{A} )$. Then, due to the inequality $E_{R}^{A|B} (\rho_{AB}) \geq S (\rho_{A}) -S(\rho_{AB} )$,
 we obtain
\begin{equation}\label{B}
E_{R} (\varrho_{m} )\geq S(\rho_{A}^{\prime\prime})-S(\varrho_m)= S [\bigtriangleup (\rho_{A} ) ]-S (\rho_{A} )=C_{R} (\rho_{A};\textbf{P}).
\end{equation}
Therefore, Eq. (\ref{S}) and Eq. (\ref{B}) can be combined to get $E_{R} (\varrho_{m} )=C_{R} (\rho_{A};\textbf{P} )$, which  completed the optimal conversion from single-partite block coherence to multipartite entanglement.

\textbf{Theorem 4.} For any multipartite quantum state $\rho_N$, the multipartite relative entropy of entanglement and block coherence are connected by the relation
\begin{eqnarray}
	E_{R} (\rho_N )\leq C_{R} (\rho_N;\textbf{P}_N ),
\end{eqnarray}
where the equality holds when the  state is the generated optimal state $\rho_{m}$ or $\varrho_{m}$.

\textbf{Proof}.---
According to the definition of multipartite relative entropy of block coherence, we have $C_{R} (\rho_N;\textbf{P}_N )=\min_{\sigma\in \mathcal{I}^N_{BI}}S (\rho_N\|\sigma )$, in which $\mathcal{I}^N_{BI}$ is the set of multipartite block-incoherent states. Since the set of multipartite block-incoherent states is a subset of multipartite separable states, we can derive that the above inequality holds. In Theorem 2, we have shown that $C_{R} (\rho_{m};\textbf{P}_m )=E_{R} (\rho_{m})$.
Next we prove that the equality also holds for $\varrho_{m}$.  Due to
\begin{equation}
	\bigtriangleup (\varrho_{m})=\sum_{i=0}^{d-1}P_{i}\rho_{A}P_{i}\otimes |0^{(i)}0^{(i)}\cdots 0^{(i)} \rangle \langle 0^{(i)}0^{(i)}\cdots 0^{(i)} |_{B_{1}B_{2}\cdots B_{n}},
\end{equation}
whose nonzero matrix elements are the same as  the ones in $\bigtriangleup (\rho_{A} )$, namely, 	 $\rho_{k^{(i)}k^{\prime(i)}}= \langle k^{(i)}0^{(i)}0^{(i)}\cdots 0^{(i)} |\bigtriangleup (\varrho_{m}) | k^{\prime(i)}0^{(i)}0^{(i)}\cdots 0^{(i)} \rangle$, we conclude that $S (\bigtriangleup [\varrho_{m} ] )=S (\bigtriangleup [\rho_{A} ] )$. Moreover, because of $S(\varrho_m)=S(\rho_A)$, the definition of relative entropy of block coherence leads to $C_{R} (\varrho_{m};\textbf{P}_r )=C_{R} (\rho_{A};\textbf{P} )$. Since $\varrho_m$ is the optimal output state, we have
\begin{equation}\label{evarrhom}
	C_{R} (\varrho_{m};\textbf{P}_r )=C_{R} (\rho_{A};\textbf{P})=E_{R} (\varrho_{m} ).
\end{equation}
  The proof is completed.
\textcolor{black}{\qed}

Note that a  result similar to Theorem 3 is also true for state $\varrho_m$. Denote $\phi_{LBICC}(\varrho_{m})$ as the state obtained by applying the LBICC operations to  $\varrho_{m}$. Since the relative entropy of block coherence is a monotone and cannot increase under the LBICC, we obtain $C_R(\varrho_m;\textbf{P}_r)\geq C_R[\phi_{LBICC}(\varrho_m);\textbf{P}_r]$. By combining Eq. (\ref{evarrhom}) with Eq. (\ref{partialtrace}), the following corollary can be obtained.

\textbf{Corollary 5.} For the optimal output state under multipartite block-incoherent operation $U_m^r$, its multipartite relative entropy of entanglement is an upper bound on the block coherence of the reduced state transformed via LBICC
\begin{equation}
	 C_{R}\left(\rho _{\beta}^{LBICC};\textbf{P}_{\beta} \right )\leq  E_{R}(\varrho_{m}),
\end{equation}
where $\rho _{\beta}^{LBICC}=\tr_{\overline{\beta}}\left[\phi_{LBICC}(\varrho_{m})\right]$ is the reduced state of multipartite state $\phi_{LBICC}(\varrho_{m})$ with $\overline{\beta}$ being the traced subsystems, and $\textbf{P}_{\beta}$ corresponds to the fixed projectors of the  remaining subsystems.

Here we give an optimal LBICC scheme that converts the multipartite entanglement of $\varrho_{m}$ to single-partite block coherence without the loss. Firstly, we apply the local block-incoherent measurement  $\{K_{j}^r=\frac{1}{\sqrt{d}}\sum_{k}e^{-i\phi _{k}^{j} }P_{j}M_{jk}P_{k}\}$ with $M_{jk}=\sum_{l}|l^{(j)}\rangle\langle l^{(k)}|$ to $B_n$.  Based on the measurement outcome $j$, the corresponding block-incoherent operation $U_j^r=\sum_{k}e^{i\phi _{k}^{j} } P_{k}$ is applied to $B_{n-1}$.  Thus the $(N+1)$-body optimal state will be transformed to  $N$-body optimal  state $\varrho_{m}^{\prime}$, which has the same form as Eq. (\ref{rhomr}) except that $|0^{(i)}\rangle\langle 0^{(j)}|_{B_n}$ is removed.  It implies that $E_{R}(\varrho_{m})=E_{R}(\varrho_{m}^{\prime})=C_{R}(\varrho_{m}^{\prime};\textbf{P}_r^{\prime})=C_{R}(\rho_{A};\textbf{P})$. Repeating the above operations until the last step in which we perform a block-incoherent measurement $\{K_j^r\}$ on the subsystem $B_1$ and then the corresponding unitary operation $U_j^r$ on  $A$, finally the quantum state of system $A$ will become $\rho_{A}^{f}=\sum_{i,j}P_{i}\rho_{A}P_{j}$. At this point,  the block coherence of the initial system $A$ is recovered, which satisfies  $C_{R}\left(\rho_{A}^{f};\textbf{P}\right)=C_{R}(\rho _A;\textbf{P})$.  Therefore,  we complete  another scheme of lossless cyclic conversion between block coherence and  multipartite entanglement, where the fixed projectors of auxiliary systems $B_1B_2\cdots B_n$ are of rank $r$.

\section{An example: resource interconversion in three four-level systems within coarse-grained quantum operations}

\begin{figure}[b]
	\centering
	\includegraphics[width=0.85\textwidth]{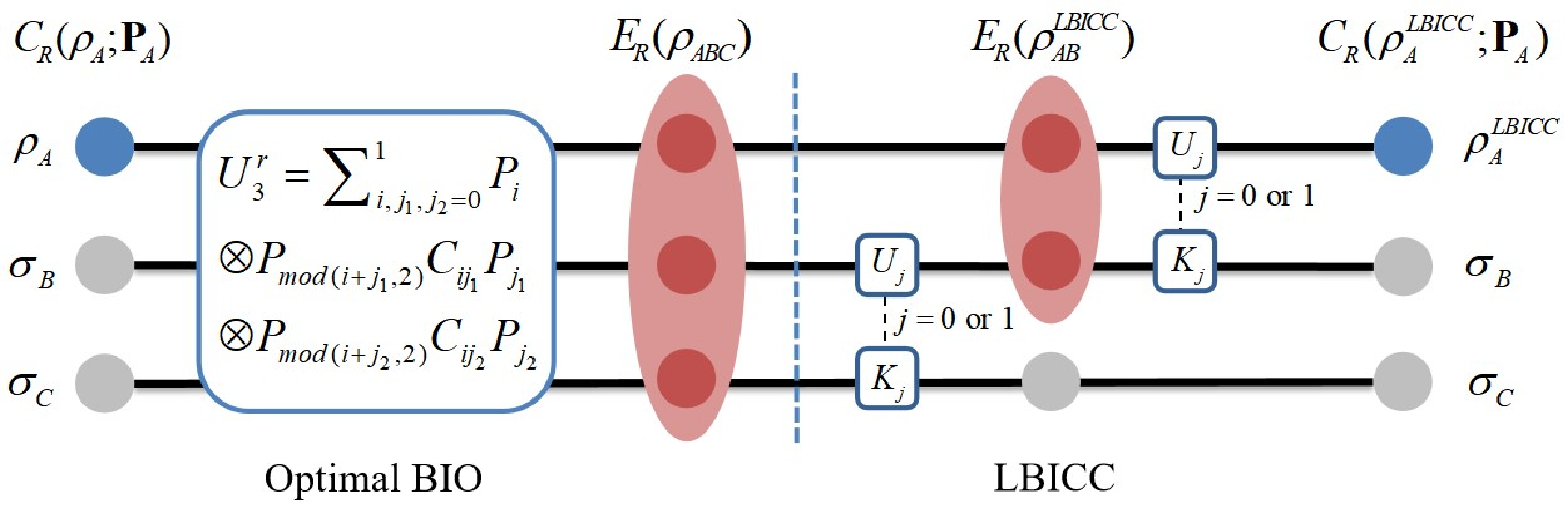}
	\caption{(Color online) A schematic diagram for the cyclic resource conversion in three four-level systems under the scenario of the coarse-grained operations, where $U_3^r$ is the optimal block-incoherent operation (BIO), and $K_j$ and $U_j$ are local block-incoherent operations assisted by classical communication.}
	\label{fig:f1}
\end{figure}

In this section, we will give an example and show how the interconversion between block coherence and multipartite entanglement can be realized in three four-level systems. As shown in figure \ref{fig:f1}, a schematic diagram is given for the cyclic resource conversion without the loss. We consider the initial single-party system in four-dimensional Hilbert space, which has the form of $\rho_A=\sum_{m,n=0}^3 \rho_{mn}|m\rangle\langle n|$. With respect to the fixed coarse-grained  projectors $\textbf{P}_A\equiv\{P_0=|0\rangle\langle 0|+|1\rangle\langle 1|, P_1=|2\rangle\langle 2|+|3\rangle\langle 3|\}$, the initial state may be block-coherent. We attach two ancillary systems  $\sigma_B=\sigma_C=|0\rangle\langle 0|$ and then apply the optimal multipartite block-incoherent operation  with respect to the  projectors $\textbf{P}_{ABC}\equiv\left\{P_{i}\otimes P_{j_{1}}\otimes P_{j_{2}}\right\}$.  In this case, the optimal tripartite block-incoherent operation has the following form:  
\begin{equation}
	U_3^{r}=\sum_{i,j_{1}, j_{2}=0}^{1}P_{i}\otimes\ P_{mod (i+j_{1},2)}C_{ij_{1}}P_{j_{1}}\otimes P_{mod (i+j_{2},2 )}C_{ij_{2}}P_{j_{2}},
\end{equation}
in which $C_{00}=|0\rangle\langle 0|+|1\rangle\langle 1|$, $C_{01}=|2\rangle\langle 2|+|3\rangle\langle 3|$, $C_{10}=|2\rangle\langle 0|+|3\rangle\langle 1|$, $C_{11}=|0\rangle\langle 2|+|1\rangle\langle 3|$.
Thus the generated tripartite quantum state can be expressed as 
\begin{eqnarray}
\rho_{ABC}&=&U_3^r(\rho_A\otimes\sigma_B\otimes\sigma_C)(U_3^r)^{\dagger}\nonumber\\
	&=&P_0\rho_A P_0\otimes|00\rangle\langle 00|+P_0\rho_A P_1\otimes|00\rangle\langle 22|\nonumber\\
	&+&P_1\rho_A P_0\otimes|22\rangle\langle 00|+P_1\rho_A P_1\otimes|22\rangle\langle 22|,
\end{eqnarray}
which satisfies $E_R(\rho_{ABC})=C_R(\rho_A;\textbf{P}_A)$.  

By means of  LBICC operations, the entanglement in tripartite  systems can be transferred to bipartite systems.  Firstly,  the observer makes measurement $\{K_j\}$ on the subsystem $C$ with the following operators:
\begin{equation}
	K_0=\frac{1}{\sqrt{2}}(P_0M_{00}P_0+P_0M_{01}P_1),\quad K_1=\frac{1}{\sqrt{2}}(P_1M_{10}P_0-P_1M_{11}P_1),
\end{equation}
in which $M_{00}=|0\rangle\langle 0|+|1\rangle\langle 1|$, $M_{01}=|0\rangle\langle 2|+|1\rangle\langle 3|$, $M_{10}=|2\rangle\langle 0|+|3\rangle\langle 1|$, $M_{11}=|2\rangle\langle 2|+|3\rangle\langle 3|$. The classical communication between $B$ and $C$ allows $B$ to perform the corresponding operation $U_0=\textbf{I}$ or $U_1=P_0-P_1$ according to the measurement outcome $j=0$ or $j=1$. After these operations, the reduced state of subsystems $AB$ becomes $\rho_{AB}^{LBICC}=P_0\rho_A P_0\otimes|0\rangle\langle 0|+P_0\rho_A P_1\otimes|0\rangle\langle 2|+P_1\rho_A P_0\otimes|2\rangle\langle 0|+P_1\rho_A P_1\otimes|2\rangle\langle 2|$, which has the same amount of  entanglement as that in tripartite systems, namely, $E_R(\rho_{ABC})=E_R(\rho_{AB}^{LBICC})$. 

Furthermore, if the measurement $\{K_j\}$ is made on the subsystem $B$ and the corresponding feedback operation $\{U_j\}$ is  performed on the subsystem $A$, then the bipartite entanglement can be converted to the initial block coherence, giving the reduced state 
\begin{equation}
\rho_A^{LBICC}=P_0\rho_A P_0+P_0\rho_A P_1+P_1\rho_A P_0+P_1\rho_A P_1,
\end{equation}
which has the same form as that of the  initial state, and satisfies  $C_R(\rho_A^{LBICC};\textbf{P}_A)=C_R(\rho_A;\textbf{P}_A)$.  We have shown that in the scenario of coarse-grained quantum operations, the block coherence and multipartite entanglement can be cyclically interconverted without the loss, which implies 
\begin{equation}
C_R(\rho_A;\textbf{P}_A)=E_R(\rho_{ABC})=E_R(\rho_{AB}^{LBICC})=C_R(\rho_A^{LBICC};\textbf{P}_A).
\end{equation}

\section{Conclusion}
In conclusion, we have established a set of rigorous quantitative relations for the interconversion between block coherence and multipartite entanglement in many-body systems. The initial single-partite block coherence  can be converted to multipartite entanglement via a multipartite block-incoherent operation. Besides, the initial block coherence also sets upper bounds on the bipartite entanglement in an arbitrary bipartition as well as the multipartite entanglement in many-body systems. In the reversed process, under the LBICC operations, the converted multipartite entanglement can be further transferred to smaller subsystems, and finally restored to block coherence in the initial single-party system. Furthermore, in the scenario of the full coarse-grained quantum operations where fine-grained projective measurements are unavailable, we have demonstrated that the lossless cyclic resource conversion between block coherence and multipartite entanglement is still realizable. As an example, we give a scheme for the cyclic resource conversion in three four-level systems.  Our results provide the advantages in the tasks of flexibly storing and utilizing quantum resources, given that observers are restricted to the measurements with different degrees of fineness.

\section*{Acknowledgments}
This work was supported by the NSF-China (Grants No. 12105074, No. 11575051 and No. 12275212), Hebei NSF (Grant No. A2021205020), Hebei 333 Talent Project (B20231005), and  Shaanxi Fundamental Science Research Project for Mathematics and Physics (Grant No. 22JSY008).

\section*{Appendix A. $\rho_m$ is a multipartite entangled state}
In the conversion from block coherence to  entanglement, Theorem 1 is also true for bipartite entanglement of arbitrary bipartition $\alpha|\overline{\alpha}$ in many-body systems $AB_1B_2\cdots B_n$.

\textbf{Corollary 6.} Applying a multipartite block-incoherent operation $\Lambda_{BI}^m$  to the initial state $\rho _{A}$ and the ancillary $N$-partite block-incoherent state $\sigma _{B_{1}B_{2} \dots B_{n}}$, the relative entropy of block coherence in $\rho _{A}$ is an upper bound on the generated bipartite relative entropy of entanglement, namely,
\begin{equation}\label{corollary5}
	C_{R} (\rho _{A};\textbf{P})\ge E_{R}^{\alpha|\overline{\alpha}} [\Lambda_{BI}^m(\rho_A\otimes\sigma _{B_{1}B_{2}\cdots B_{n}} ) ],
\end{equation}
where $\textbf{P}=\{P^A_{i}\}$,  $\Lambda_{BI}^m$ is an $(N+1)$-partite block-incoherent operation with respect to the general projective measurement $\{P_i^A\otimes P^{B_1}_{j_1}\otimes P^{B_2}_{j_2}\otimes\cdots\otimes P^{B_n}_{j_n}\}$, and  $E_R^{\alpha|\overline{\alpha}}$ is the bipartite relative entropy of entanglement in any bipartition $\alpha|\overline{\alpha}$ with $\alpha\cup\overline{\alpha}=AB_1B_2\cdots B_n$.

\textbf{Proof}.---
Letting $\sigma_{A}$ be  the closest block-incoherent state to $\rho_{A}$, then according to the definition of relative entropy of block coherence,  we have
\begin{eqnarray}
	C_{R}(\rho_{A};\textbf{P})&=&S(\rho_{A}\|\sigma_{A}) \nonumber\\
	&=&S\left ( \rho _{A}\otimes \sigma _{B_{1}B_{2} \dots B_{n}} \parallel \sigma _{A}\otimes \sigma _{B_{1}B_{2} \dots B_{n}}  \right )\nonumber\\
	&\ge& S\left [ \Lambda_{BI}^m \left ( \rho _{A}\otimes \sigma _{B_{1}B_{2} \dots B_{n}} \right )\parallel \Lambda_{BI}^m\left ( \sigma _{A} \otimes \sigma _{B_{1}B_{2} \dots B_{n}} \right ) \right ]\nonumber\\
	&\ge& E_{R}^{\alpha|\overline{\alpha}}\left [ \Lambda_{BI}^m\left ( \rho _{A}\otimes \sigma _{B_{1}B_{2} \dots B_{n}}\right ) \right ],
\end{eqnarray}
where the additive and
contractive properties of relative entropy are used in the second  and  third lines, and in the last step, since the quantum state $\Lambda_{BI}^m \left ( \sigma _{A}\otimes \sigma _{B_{1}B_{2} \dots B_{n}} \right )$ is  multipartite block-incoherent, which is
not only fully separable, but also bipartite separable in an arbitrary bipartition $\alpha|\overline{\alpha}$   such as $A|B_1B_2\cdots B_n, AB_1|B_2\cdots B_n,  \cdots$ , and then the last inequality can be satisfied for the corresponding bipartite entanglement.
\textcolor{black}{\qed}

With reference to fine-grained projective measurement  in the auxiliary systems, i.e., $\{P_{i}\otimes|j_{1}\rangle\langle j_{1}|\otimes\cdots\otimes|j_{n}\rangle\langle j_{n}|\} $, the optimal multipartite block-incoherent operation $U_m$ performed on $\rho_A$ and $|00\cdots0\rangle\langle00\cdots0|_{B_{1}B_{2}\cdots B_{n}}$ gives the output state
\begin{equation}
	\rho_{m}=\sum_{i,j=0}^{d-1}P_{i}\rho_{A}P_{j}\otimes\left|ii\cdots i\right\rangle\left\langle jj\cdots j\right|_{B_{1}B_{2}\cdots B_{n}}.
\end{equation}
It is noted that $\rho_m$ is similar to the
maximally correlated state \cite{Rains99pra,Rains01,Vidal02prl}, apart from the form of subsystem $A$.
In an arbitrary bipartition $\alpha|\overline{\alpha}$, the reduced state can be obtained by  $\rho_{\alpha}=\tr_{\overline{\alpha}}\rho_m=\sum_{i}P_{i}\rho_{A}P_{i}\otimes\left|ii\cdots i\right\rangle_{\alpha/A}\left\langle ii\cdots i\right|$, in which $\alpha/A$ represents the subsystem $\alpha$ with $A$ being removed, and $\alpha$ could take $A,AB_1,AB_2,AB_1B_2,$ and so on.
Since the matrix elements of  $\bigtriangleup(\rho_{A})$  have the form of $\rho_{k^{(i)}l^{(i)}}= \langle k^{(i)} |P_{i} \rho _{A}P_{i} | l^{({i})  }  \rangle= \langle k^{(i)}ii\cdots i  |\rho_{\alpha} | l^{ (i)  }ii\cdots i \rangle$, which are  embedded in the matrix of $\rho_{\alpha}$, we obtain $S(\rho_{\alpha})=S[\bigtriangleup(\rho_{A})]$.  Due to $S\left(\rho_{m}\right)=S\left(\rho_{A}\right)$ and $E^{A:B}_{R}\left(\rho_{AB}\right)\geq S\left(\rho_{A}\right)-S\left(\rho_{AB}\right)$ \cite{Plenio},
 we obtain
\begin{eqnarray}
	E_R^{\alpha|\overline{\alpha}}(\rho_m) &\geq& S(\rho_{\alpha})-S(\rho_m) \nonumber\\
	&=& S\left[\bigtriangleup\left(\rho_{A}\right)\right]-S\left(\rho_{A}\right)\nonumber\\
	&=&C_{R}\left(\rho_{A};\textbf{P}\right).
\end{eqnarray}
 Furthermore, according to Corollary 6, we get $E_{R}^{\alpha|\overline{\alpha}}(\rho_{m})\leq C_{R}(\rho_{A};\textbf{P})$. Therefore, $E_R^{\alpha|\overline{\alpha}}(\rho_m)=C_{R}\left(\rho_{A};\textbf{P}\right)$. That is to say, if the initial state $\rho_A$ is block-coherent, the bipartite relative entropy of entanglement of  state $\rho_m$ for any bipartition is equal to the initial block coherence of $\rho_A$. Since the bipartite entanglement of $\rho_m$ is not zero for  any bipartition,  we can say that $\rho_m$ is a multipartite entangled state.

\section*{Appendix B. $U_m^{r}$ is an optimal multipartite block-incoherent operation}
In this appendix, we  show that $U_m^{r}$ is an optimal multipartite block-incoherent operation  when $\textbf{P}_r=\left\{P_{i}\otimes P_{j_{1}}\otimes\cdots\otimes P_{j_{n}}\right\}$ in which each projector has the same rank $r$ with the form of
$P_{i}=\sum_{k=0}^{r-1}|k^{(i)}\rangle\langle k^{(i)}|$. Here $\{|k^{(i)}\rangle\}_{k}$ represents a basis of subspace given by $P_{i}$.
The  form of $U_m^{r}$ reads
\begin{equation}
	U_m^r=\sum_{i,j_{1},\cdots j_{n}=0}^{d-1}P_{i}\otimes\ P_{mod (i+j_{1},d)}C_{ij_{1}}P_{j_{1}}\otimes\cdots\otimes\ P_{mod (i+j_{n},d )}C_{ij_{n}}P_{j_{n}},
\end{equation}
where $C_{ij}=\sum_{n=0}^{r-1} |n^{mod (i+j,d )} \rangle \langle n^{ (j )} |$.

For a multipartite system, choosing the fixed reference projectors to be $\textbf{P}_r$, a multipartite block-incoherent state can be defined as
\begin{equation}\label{blockincoherent}
	\sigma_{AB_{1}\cdots B_{n}}=\sum_s p_s \sigma_s^{A}\otimes \sigma_s^{B_1}\otimes\cdots \otimes\sigma_s^{B_n},
\end{equation}
where $p_{s}$ are probabilities, $\sigma_s^{A}$ is a block-incoherent state on the subsystem $A$, namely, $\sigma_s^{A}=\sum_{i}P_{i}\rho_s^{A}P_{i}$ with $\rho_s^{A}$ being any state in the Hilbert space of subsystem $A$, and so do $\sigma_s^{B_k}(k=1,\cdots,n)$.
More specifically, $\sigma_s^{A}$ can also be written in the form of $\sigma_s^{A}=\sum_{i}\sum_{k,k'}\rho_{k^{(i)}k'^{(i)}}|k^{(i)}\rangle \langle k'^{(i)}|$, in which $\rho_{k^{(i)}k'^{(i)}}$ is the  matrix element determined by the bases $|k^{(i)}\rangle$ and $| k'^{(i)}\rangle$. Similarly,  assuming that $\{|l_{1}^{(j_{1})}\rangle\},\cdots,\{|l_{n}^{(j_{n})}\rangle\}$ are the corresponding  basis of the subspaces given by  $P_{j_{1}},\cdots,P_{j_{n}}$ respectively, thus we can also rewrite $\sigma _{s}^{B_{1} }=\sum _{j_{1},l_{1} ,l_{1}' }\rho _{l_{1}^{(j_{1}) }l_{1}'^{(j_{1} )}}|l_{1}^{(j_{1}) }\rangle\langle l_{1}'^{(j_{1} )}|,\cdots,\sigma _{s}^{B_{n} }=\sum _{j_{n},l_{n} ,l_{n}' }\rho _{l_{n}^{(j_{n}) }l_{n}'^{(j_{n} )}}|l_{n}^{(j_{n}) }\rangle\langle l_{n}'^{(j_{n} )}|$.

The block-incoherent operations can be expressed by Kraus operators $\{K_{l}\}$ which satisfy the conditions $\sum_{l}K_{l}^{\dagger}K_{l}=\textbf{I}$ and $K_{l}\mathcal{I}_{BI}^{N}K_{l}^{\dagger}\subseteq\mathcal{I}_{BI}^{N}$ with $\mathcal{I}_{BI}^{N}$ being now the set of $(N+1)$-partite block-incoherent states.
To prove that $U_m^{r}$ is a multipartite block-incoherent operation, we should verify $U_m^{r}(U_m^{r})^{\dagger}=\textbf{I}$ and $U_m^r \sigma_{AB_{1}\cdots B_{n}} (U_m^r)^{\dagger}\subseteq\mathcal{I}_{BI}^{N}$.
The proof process goes as follows.

First, we prove $U_m^r$ is an unitary operation.
\begin{eqnarray}\label{1}
&&U_m^{r}(U_m^{r})^{\dagger} \nonumber\\
&=& (\sum_{i,j_{1},\cdots,j_{n}}P_{i}\otimes P_{mod (i+j_{1},d )}C_{i,j_{1}}P_{j_{1}}\otimes\cdots\otimes P_{mod (i+j_{n},d )}C_{i,j_{n}}P_{j_{n}} )\nonumber\\
&& (\sum_{i',j_{1}',\cdots,j_{n}'}P_{i'}\otimes P_{mod (i'+j_{1}',d )}C_{i',j_{1}'}P_{j_{1}'}\otimes\cdots\otimes P_{mod (i'+j_{n}',d )}C_{i',j_{n}'}P_{j_{n}'} )^{\dagger} \nonumber\\ &=&\sum_{i,j_{1},\cdots,j_{n}}P_{i}\otimes P_{mod (i+j_{1},d )}C_{i,j_{1}}P_{j_{1}}C_{i,j_{1}}^{\dagger}P_{mod (i+j_{1},d )}
\otimes\cdots\otimes\nonumber\\
&& P_{mod (i+j_{n},d )}C_{i,j_{n}}P_{j_{n}}C_{i,j_{n}}^{\dagger }P_{mod (i+j_{n},d )},
\end{eqnarray}
where
\begin{eqnarray}\label{2}
C_{i,j_{1}}P_{j_{1}}C_{i,j_{1}}^{\dagger}&=&\sum_{n} |n^{mod (i+j_{1},d )} \rangle \langle n^{(j_{1})} |\sum_{l_{1}} |l_{1}^{ (j_{1} )} \rangle \langle l_{1}^{ (j_{1} )} |\sum_{n'} |n'^{ (j_{1} )} \rangle \langle n'^{mod (i+j_{1} ,d)} |\nonumber\\
&=&\sum_{n,l_{1},n'} |n^{mod (i+j_{1},d )} \rangle \langle n'^{mod (i+j_{1} ,d)} |\delta_{n,l_{1}}\delta_{l_{1},n'}\nonumber\\
&=&P_{mod (i+j_{1},d )}.
\end{eqnarray}
Similarly, $C_{i,j_{2}}P_{j_{2}}C_{i,j_{2}}^{\dagger}=P_{mod (i+j_{2},d )},\cdots, C_{i,j_{n}}P_{j_{n}}C_{i,j_{n}}^{\dagger}=P_{mod (i+j_{n},d )}$, which  implements the permutation of projectors. It is worth mentioning that $C_{ij}=\sum_{n=0}^{r-1} |n^{mod (i+j,d )} \rangle \langle n^{ (j )} |$ is a mapping of bases with the same label $n$ between projectors $P_{i}$ and $P_{j}$. Actually the matrix $C$ can be constructed in different ways, as long as it realizes one-to-one mapping between the bases of $P_{i}$ and $P_{j}$.  Due to Eqs. (\ref{1}) and  (\ref{2}), we obtain
\begin{equation}
U_m^{r}(U_m^{r})^{\dagger}=\sum_{i,j_{1},\cdots,j_{n}=0}^{d-1}P_{i}\otimes P_{mod (i+j_{1},d )}\otimes\cdots\otimes P_{mod (i+j_{n},d )}=\textbf{I}.
\end{equation}

Next, we apply $U_m^r$ to an arbitrary  multipartite block-incoherent state which has been defined in Eq. (\ref{blockincoherent}), and then
\begin{eqnarray}
&&U_m^{r}\sigma_{AB_{1}\cdots B_{n}}(U_m^{r})^{\dagger}\nonumber\\
&=& (\sum_{i,j_{1},\cdots,j_{n}}P_{i}\otimes P_{mod (i+j_{1},d )}C_{i,j_{1}}P_{j_{1}}\otimes\cdots\otimes P_{mod (i+j_{n},d )}C_{i,j_{n}}P_{j_{n}} )\nonumber\\
&&\sum_s p_s \sigma_s^{A}\otimes \sigma_s^{B_1}\otimes\cdots \otimes\sigma_s^{B_n}\nonumber\\
&& (\sum_{i',j_{1}',\cdots,j_{n}'}P_{i'}\otimes P_{mod (i'+j_{1}',d )}C_{i',j_{1}'}P_{j_{1}'}\otimes\cdots\otimes P_{mod (i'+j_{n}',d )}C_{i',j_{n}'}P_{j_{n}'} )^{\dagger} \nonumber\\
&=&\sum_s p_s \sum_{i,j_{1},\cdots,j_{n}}\sum_{i',j_{1}',\cdots,j_{n}'}P_{i}\sigma_s^{A}P_{i'}
\otimes P_{mod (i+j_{1},d )}C_{i,j_{1}}P_{j_{1}}\sigma_s^{B_1}P_{j_{1}'}C_{i',j_{1}'}^{\dagger}P_{mod (i'+j_{1}',d )}\nonumber\\
&&\otimes\cdots\otimes P_{mod (i+j_{n},d )}C_{i,j_{n}}P_{j_{n}}\sigma_s^{B_n}P_{j_{n}'}C_{i',j_{n}'}^{\dagger}P_{mod (i'+j_{n}',d )}.
\end{eqnarray}
Substituting $\sigma_s^{A}=\sum_{i}\sum_{k,k'}\rho_{k^{(i)}k'^{(i)}}|k^{(i)}\rangle \langle k'^{(i)}|,\sigma _{s}^{B_{1} }=\sum _{j_{1},l_{1} ,l_{1}' }\rho _{l_{1}^{(j_{1}) }l_{1}'^{(j_{1} )}}|l_{1}^{(j_{1}) }\rangle\langle l_{1}'^{(j_{1} )}|,\cdots$  into the above equation, then
\begin{eqnarray}
	&&U_m^{r}\sigma_{AB_{1}\cdots B_{n}}(U_m^{r})^{\dagger}\nonumber\\
&=&\sum_{s}p_{s}\sum_{i,j_{1},\cdots,j_{n}} \sum_{k,k'}\rho_{k^{
(i)}k'^{(i)}} |k^{(i)}\rangle \langle k'^{ (i )} |\otimes\sum_{l_{1},l_{1}'}\rho_{l_{1}^{ (j_{1} )}l_{1}'^{ (j_{1} )}} |l_{1}^{(mod[i+j_{1},d] )} \rangle \langle l_{1}'^{ (mod[i+j_{1},d] )} |\nonumber\\
&&\otimes\cdots\otimes\sum_{l_n,l_{n}'}\rho_{l_{n}^{ (j_{n} )}l_{n}'^{ (j_{n})}} |l_{n}^{ (mod[i+j_{n},d] )} \rangle \langle l_{n}'^{ (mod[i+j_{n},d] )} |\nonumber\\
&=&\sum_s p_s \sigma_s'^{A}\otimes \sigma_s'^{B_1}\otimes\cdots \otimes\sigma_s'^{B_n}\nonumber\\
&=&\sigma_{AB_{1}\cdots B_{n}}'\subseteq\mathcal{I}_{BI}^{N}.
\end{eqnarray}
Therefore, $U_m^{r}$ is a multipartite block-incoherent operation.

Applying $U_m^{r}$ to the initial state $\rho_A$ and its ancillas $|0^{(0)}0^{(0)}\cdots0^{(0)}\rangle\langle 0^{(0)}0^{(0)}\cdots0^{(0)}|_{B_1B_2\cdots B_n}$, the generated state is
\begin{eqnarray}\label{varrho}
	\varrho_{m}&=& U_m^r (\rho_{A}\otimes |0^{(0)}0^{(0)}\cdots0^{(0)}\rangle\langle 0^{(0)}0^{(0)}\cdots0^{(0)}|)(U_m^r)^{\dagger}\nonumber\\
	&=&\sum_{i,j=0}^{d-1} P_{i}\rho_{A}P_{j}\otimes |0^{(i)}0^{(i)}\cdots 0^{(i)} \rangle \langle 0^{(j)}0^{(j)}\cdots0^{(j)} |_{B_1B_2\cdots B_n}.
\end{eqnarray}
Since we have proved $E_{R} (\varrho_{m} )=C_{R} (\rho_{A};\textbf{P} )$ in the main text, we can conclude that $U_m^{r}$ is an optimal multipartite block-incoherent operation.

Furthermore, $\varrho_m$ is also a multipartite entangled state. According to
Corollary 6, the bipartite relative entropy of entanglement of $\varrho_m$ for an arbitrary bipartition $\beta|\overline{\beta}$ is not larger than the initial block coherence, namely, $E_{R}^{\beta|\overline{\beta}}(\varrho_{m})\leq C_{R}(\rho_{A};\textbf{P})$. The reduced state can be obtained by  $\rho_{\beta}=\tr_{\overline{\beta}}\varrho_m=\sum_{i}P_{i}\rho_{A}P_{i}\otimes|0^{(i)}0^{(i)}\cdots 0^{(i)}\rangle_{\beta/A}\langle 0^{(i)}0^{(i)}\cdots 0^{(i)}|$, in which $\beta/A$ is the subsystem $\beta$ with $A$ being removed, and $\beta$ could take $A,AB_1,AB_1B_2,\cdots$.
Similarly, the nonzero matrix elements of $\bigtriangleup(\rho_{A})$   are  the same as those of $\rho_{\beta}$, which means $S(\rho_{\beta})=S[\bigtriangleup(\rho_{A})]$.  Due to $S\left(\varrho_{m}\right)=S\left(\rho_{A}\right)$ and $E^{\beta|\overline{\beta}}_{R}\left(\varrho_{m}\right)\geq S\left(\rho_{\beta}\right)-S\left(\varrho_{m}\right)$, we have
$E_R^{\beta|\overline{\beta}}(\varrho_m) \geq C_{R}\left(\rho_{A};\textbf{P}\right)$.
Therefore, $E_R^{\beta|\overline{\beta}}(\varrho_m)=C_{R}\left(\rho_{A};\textbf{P}\right)$ for any bipartition $\beta|\overline{\beta}$, which means that $\varrho_m$ is  multipartite entangled.

\end{document}